# A Decision Tree Approach to Classify Web Services using Quality Parameters

Shilpa Sonawani[1] and  Debajyoti Mukhopadhyay[2]

[1]Department of Computer Engineering
[2]Department of Information Technology
Maharashtra Institute of Technology
Pune 411038, India
{shilpasonawani,debajyoti.mukhopadhyay}@gmail.com

**Abstract:** With the increase in the number of web services, many web services are available on internet providing the same functionality, making it difficult to choose the best one, fulfilling user's all requirements. This problem can be solved by considering the quality of web services to distinguish functionally similar web services. Nine different quality parameters are considered. Web services can be classified and ranked using decision tree approach since they do not require long training period and can be easily interpreted. Various decision tree and rules approaches available are applied and tested to find the optimal decision method to correctly classify functionally similar web services considering their quality parameters.

**Keywords:** Web Services, QoS, Decision Tree Classification
.

## 1    Introduction

In today's world of internet, web and ecommerce applications have become very popular because of the heterogeneous infrastructure of internet in a distributed environment. These distributed applications can be accessed through the web service technology. Starting a new business has become easier just by reusing, extending or merging existing web services with the other services and providing customers the best possible service at one place instead of searching for different services for different purposes.

The web service is an interface that has collection of number of operations and messages, converting user application to web application. They use XML to code and decode data and SOAP to transport it. That makes the applications platform independent which makes it very popular on internet. Web services are composed of following platform elements, SOAP(simple object access protocol), UDDI(universal description, discovery and information) and WSDL(web services description language). Web services are registered by service providers in UDDI which is keeping information about the company details, web service description and their functions and supported features. The requesters are finding their service of interest through UDDI Business Registries (UBRs). UBRs are not efficient enough to find the relevant web service of interest. Currently it has keyword based search operations. The UDDI specification does not include QoS as part of its publication [3] or request handling. Currently no such algorithm or model is used to evaluate and determine similar web services according to their Qos criteria.

Quality of a service is a measure of how well it is serving the request of the requester. With the increase in the number of web services, many web services providing the same functionality making it very difficult to choose the best web service fulfilling all the requirements. So quality of web services has become an important issue to distinguish functionally similar web services by ranking them.

## 2    Related Work

To solve this problem some work has been implemented for enhancing the search techniques for When there are many number of web services available for the same functionality it is very difficult to choose the one which is most relevant, their non functional properties can be used for choosing the best one. Current UDDI does not include QoS parameters as part of their search.

To solve this problem some work has been implemented for enhancing the search techniques for UBR by including the QoS information within the messages [4].  In another approach, Web service relevancy function (WSRF) [1] is used for measuring the relevancy ranking of a web services based on their QoS metrics and client preferences. They proposed a WSRB framework which uses a web service crawler Engine (WSCE) that actively crawls UBRs for measuring the QoS information for the web services and the information is stored in Web service Storage (WSS). WSRB enables clients choose and manage their search criteria through graphical interface. WsrF is using QoS parameters for computing relevancy ranking. The different QoS parameters considered are response time, throughput, availability, accessibility, interoperability analysis and cost of service. Other approach [2] proposed a web service QoS manager which is a Trusted service Broker containing QoS parameters for the web services.

Another ranking model [5] is proposed to rank and recommend a web service using artificial neural network by measuring QoS parameters. It proposes a principal component analysis (PCA) method for initial attribute weight then gives training algorithm for weight adjusting based on neural network. A backpropagation based neural network [6] has been used to discover the high quality web service.  Although neural networks take long time for training for large datasets but it was shown that[5,6] at a starting point neural networks could  be used to discover and rank the web service and there is room for use of some other type of neural networks such as Fuzzy ARTs, ARTMAPs and Self organized Map.  Naïve based Bayesian network [7] can also be used for classification of the services.

Many QoS parameters enabled discovery mechanisms have been proposed but still it is a challenge to include them because it is an extra overhead and controlling, managing it is also challenging. Ensuring the validity of the information collected and timely updating of the information is also need to be considered.

## 3    Classification using Decision Tree

Classification techniques in data mining can predict categorical class labels (discrete or nominal). They classify data by constructing a model based on the training set provided; the values (class labels) used in a classifying attribute and use it in classifying new data.

Decision tree approach is a supervised classification. It has simple structure with non terminal nodes representing tests on one or more attributes and terminal nodes reflect decision outcomes or class labels. In order to classify unknown sample, its values are tested against decision tree. Decision trees can be easily converted into decision rules. When the trees are built many of the branches may reflect may reflect noise or outliers in the training data. Tree Pruning can identify such branches and remove them which improve the accuracy of the classification results on the unseen data.

The basic algorithm of decision tree is using a greedy concept, the steps are given below
1. Tree is constructed in a top-down recursive divide-and-conquer manner
2. All the training examples are at the start in the beginning as a single node
3. If all samples of the same class, then the node becomes a leaf and is labeled with a class
4. Attribute that will best separate the samples into individual classes is selected using Entropy based measure known as information gain as heuristic. Attributes are categorized, that is, discrete valued. Continuous valued are discretized
5. On selected attributes the data is recursively partitioned

Conditions for stopping partitioning
1. All the samples for the given node belong to the same class
2. No attributes are remaining for partitioning
3. No samples left for partitioning

Many different measures are used to select the best attribute to separate the samples. Information gain is one of the measures; an attribute with highest information gain is selected. ID3 uses information gain. Let *pi* be be the probability that any tuple in D belongs to class Ci, estimated by |Ci,D|/|D|. Information gain is defined as the difference between the information based on the proportion of classes and the information obtained after partitioning on A. That is [11]

$$Gain(A) = Info(D) - Info_A(D)$$

*Info(D)* is the Expected information (entropy) required to classify a tuple in D which is calculated as

$$Info(D) = -\sum_{i=1}^{m} p_i \log_2(p_i)$$

*InfoA(D)* is Information required after using A to split D into v partitions to classify D which as calculated as

$$Info_A(D) = \sum_{j=1}^{v} \frac{|D_j|}{|D|} \times I(D_j)$$

It is found that Information gain tend to prefer attributes with many values. Another measure, gain ratio used in C4.5 used to overcome the problem. The attribute with the maximum gain ratio is selected as the splitting attribute which is calculated as

$$GainRatio(A) = Gain(A)/SplitInfo(A)$$

Where

$$SplitInfo_A(D) = -\sum_{j=1}^{v} \frac{|D_j|}{|D|} \times \log_2(\frac{|D_j|}{|D|})$$

Gini Index can also be used which produces equal-sized partitions bringing purity in both partitions. It is used in CART. The attribute with maximum reduction in impurity is selected as the splitting attribute. If a data set D contains examples from n classes, gini index, gini(D) is defined as

$$gini(D) = 1 - \sum_{j=1}^{n} p_j^2$$

Considering binary split, a weighted sum of the impurity of each resulting partition is computed as

$$gini_A(D) = \frac{|D_1|}{|D|} gini(D_1) + \frac{|D_2|}{|D|} gini(D_2)$$

The reduction in the impurity on due to binary split would be

$$Gini(A) = Gini(D) - Gini_A(D)$$

Many alternatives have been proposed, Chi square contingency table statistics, G-statistic, CHAID, C-SEP, MDL (Minimal Description Length) principle, Multivariate splits.

Unlike Neural Networks, decision tree methods are able to identify independent variables through the built tree and basic functions when many potential variables are considered [8]. When the dataset is huge they can save lots of modeling time since they do not need a long training process. One very strong advantage of decision tree is that the resulting classification model can be easily interpreted. They not only point out which variables are important in classifying objects, but also indicate that a particular object belongs to a specific class when the built rules are satisfied[8].The advantages of decision tree methods are listed as follows:
1.  Decision trees are simple and easy to understand
2.  Decision trees are easily converted to a set of production rules
3.  Decision trees can classify both categorical and numerical data, but categorical attribute as output
4.  No prior assumptions are required for the data.

Although decision tree algorithms also have their disadvantages like multiple output attributes are not allowed, slight variations in the training dataset can cause different attribute selections at each choice point within the tree. The attribute selection can affect all descendent trees. Trees created from numeric datasets can be complex since could be binary. If the tree is not pruned, it could be large. But they are effective tools in handling forecasting and classification problems ((McGlynn, et al. 2004; Zhang & Zhao, 2007)).

## 4  Decision Tree Algorithms

Decision trees considered here for study are from WEKA (The Waikato Environment for Knowledge Analysis). WEKA is a data mining tool for data analysis and contains implementations of data pre-processing, classification, clustering, association rules, and visualization by different algorithms.

The different decision tree algorithms included in WEKA are briefly explained below:

### 4.1   REPTree:

It is a fast decision tree learner building a decision/regression tree using information gain as the splitting criterion, and using reduced error pruning prunes it (with backfitting). It sorts values for numeric attributes once. Missing values are dealt with using C4.5's method of using fractional instances. REPTree considers all the attributes.

### 4.2   RandomTree:

It considers a set of K randomly chosen attributes to split on at each node. Random here means that each tree in the set of trees has an equal chance of being sampled and hence making uniform distribution of trees. Random trees can be generated efficiently and the combination of large sets of random trees generally leads to accurate models. Random tree models have been extensively developed in the field of Machine Learning in the recent years. It performs no pruning.

### 4.3   Random Forest:

The random forest machine learner, is a meta-learner; meaning consisting of many individual learners (trees). The random forest uses multiple random trees classifications to votes on an overall classification for the given set of inputs. Many features of the random forest algorithm have yet to be implemented into this software. Tree cannot be visualized in the explorer.

**4.4    J48:**

It is a new version of an earlier very popular algorithm C4.5 Decision trees developed by J. Ross Quinlan. It is providing variety of options which generates an unpruned or pruned C4.5 decision tree. The C4.5 algorithm generates a classification-decision tree for the given data-set by recursive partitioning of data. The basic algorithm recursively classifies until each leaf is pure that is the data has been categorized perfectly as possible ensuring maximum accuracy on the training data.

**4.5    DecisionStump**:

It is a one level decision tree. It has one internal node which is connected directly to the terminating nodes. A single root node decides how to classify inputs based on a single feature. Each leaf represents possible feature value, the class label that should be assigned to inputs whose features have that value. For using this method, one must decide the feature and build the tree. It is the simplest method by which decision stumps can be build for each possible feature and which feature is giving highest accuracy on the training data can be checked.

**4.6    LMT:**

Logistic Model Trees use logistic regression functions at the leaves. This method can deal with missing values, binary and multi-class variables, numeric and nominal attributes. It generates small and accurate trees. It uses CART pruning technique. It does not require any tuning parameters. It is often more accurate than C4.4 decision trees and standalone logistic regression [9]. LMT produces a single tree containing binary splits on numeric attributes, multiway splits on nominal ones and logistic regression models at the leaves. It also ensures that only relevant attributes are included in the latter.

**4.7    Decision Table :**

This is the class for building and using a simple decision table majority classifier (Ron Kohavi). Decision Table employs the wrapper method to find a good subset of attributes for inclusion in the table using a best-first search. By eliminating attributes that contribute little or nothing to a model of the dataset, the algorithm reduces the likelihood of over-fitting and creates a smaller and condensed decision table.

**4.8    ZeroR:**

It predicts the majority class in the training data for problems with a categorical class value, and the average class value for numeric prediction problems. It predicts the mean for numeric value & mode for nominal class. It is useful for assuming a baseline performance for comparing with other learning schemes.

**4.7    OneR:**

It produces very simple rules based on a single attribute. It produces one rule for each attribute in training dataset then select the one rule as a result by choosing the rule which has the lowest error rate. In WEKA oneR algorithm considers the rule with the highest number of correct instances not the lowest error rate and when the error rates are same the it is not randomly selecting a rule. The rules produced might not be accurate compared to state of the art classification methods but they are simple and easy to interpret. This algorithm also treats all numerically-valued attributes as continuous and divides the range of values into several disjoint intervals introducing the risk of 'overfitting' for continuously-valued attributes.

    **a.    PART:**

This algorithm generates ordered set of rules called decision lists. New data is compared with each rule in the list and the data is assigned the category of the rule to which it is best matching. It is a combination of JRip and C4.5.

### 4.9 JRip

This class implements a propositional rule learner. This algorithm is developed by William W. Cohen as an optimized version of IREP. It performs Repeated Incremental Pruning to Produce Error Reduction (RIPPER). JRip is a bottom–up method learns rules by treating particular judgment of the examples in the training data as a class and finding the set of rules covering all the members of the class. Cross-validation and minimum-description length techniques are used to prevent overfitting.

## 5 Mechanism

A mechanism can be developed on the server side where the WSCE(Web Service Crawler Engine) is collecting quality information for the web services. The decision tree classification model is built on the dataset. Many times the users are not interested in all the quality parameters which sometimes make trade off, if the user is interested in web services with high response time, latency and low successability, he/she can choose the quality parameters at the time of giving search query.

On the server side multidimensional cubes can be created for various combinations of quality parameters. The classification models are built in advance for the various data cubes. Multidimensional cubes are using the sparse matrix technology which is making the storage of multidimensional data cubes highly efficient. Since the classification models are pre built for various quality parameters, whenever the user is posting a query, the functionally similar web services with selected or all quality parameters can give fast response. The decision rule algorithms used here of generating classification model simple, easy to understand and require less time for rules generation and can work on multiple attributes making the whole mechanism easy for management.

## 6 Experimental Results

Publicly available Quality of web services data is used for classification of web services using decision tree approach. QWS dataset is [10] composed of 2507 real web services with their QWS measurements measured on web. The web services information is collected using the Web Service Crawler Engine (WSCE) from public sources, UBR, service portals and search engines and quality measurements were conducted using Quality Web Service Manager. All the web services are measured for nine parameters which are listed in Table 1 along with parameter description and units of measurements. The sample experimental dataset consists of 364 functionally similar web services for a query with the keyword "sms" available on web in demo section [10].

The statistical data collected so far can be further analyzed using WEKA, a knowledge analyzer. Weka is a open source software which is used for data mining task having collection of machine learning algorithms. Classification trees in WEKA can be used to correctly classify web services in classes which can be then used to priorities the web services. Table 2 shows the classification result observations of the various classification tree techniques in WEKA, showing accuracy of the method and time to build the model in seconds obtained using 10-fold cross validation. It is observed that Decision Table (100)% and JRip(100%) methods are performing excellent with JRip optimal method taking the less time to build the model with good accuracy.

**Table 1.** QwsParameters

| ID | Parameter | Description | Unis |
|---|---|---|---|
| 1 | Response Time | Time to send a request and receive a response | ms |
| 2 | Availability | Successful invocations/total invocations | % |
| 3 | Throughput | Total Number of invocations /period of time | Invokes/s |
| 4 | Successability | Number of response/number of request messages | % |
| 5 | Reliability | Ratio: number of error messages/total messages | % |
| 6 | Compliance | Extent a WSDL follows a specification | % |
| 7 | Best Practices | Extent a WSDL follows a specification | % |
| 8 | Latency | Time to process a given request | Ms |
| 9 | Documentation | Measure of documentation in WSDL | % |

**Table 2**: Classification result observations:

| Tree | Accuracy(%) | Time/Sec |
|---|---|---|
| DecisionStump | 60.43 | 0.02 |
| J48 | 99.73 | 0.05 |
| LMT | 99.73 | 2.45 |
| Random Forest | 98.35 | 0.06 |
| RandomTree | 96.43 | 0.03 |
| REPTree | 99.73 | 0.03 |
| Decision Table | 100 | 0.28 |
| JRip | 100 | 0.06 |
| OneR | 99.73 | 0 |
| PART | 99.11 | 0.05 |
| ZeroR | 32.967 | 0 |

**Table 3**: Classification Instances observations

| Tree | Correctly Classified Instances | Incorrectly Classified Instances |
|---|---|---|
| DecisionStump | 220 | 144 |
| J48 | 363 | 1 |
| LMT | 363 | 1 |
| Random Forest | 358 | 6 |
| RandomTree | 351 | 13 |
| REPTree | 363 | 1 |
| Decision Table | 364 | 0 |
| JRip | 364 | 0 |
| OneR | 363 | 1 |
| PART | 363 | 1 |
| ZeroR | 120 | 244 |

Table 4: Classification result analysis

| Tree | Mean absolute error | Root mean squared error | Relative absolute error | Root relative squared error |
|---|---|---|---|---|
| Decision Stump | 0.2322 | 0.3407 | 64.1853 | 80.1364 |
| J48 | 0.0014 | 0.0371 | 0.3798 | 8.7168 |
| LMT | 0.0122 | 0.059 | 3.3705 | 13.8764 |
| Random Forest | 0.0337 | 0.0927 | 9.304 | 21.7919 |
| Random | 0.0179 | 0.1336 | 4.9368 | 31.4287 |
| REPTree | 0.0018 | 0.0372 | 0.511 | 8.7567 |
| Decision Table | 0.0176 | 0.0219 | 4.8545 | 5.1572 |
| Jrip | 0 | 0 | 0 | 0 |
| OneR | 0.0014 | 0.0371 | 0.3798 | 8.7168 |
| PART | 0.0014 | 0.0371 | 0.3798 | 8.7168 |
| ZeroR | 0.3617 | 0.4252 | 100% | 100% |

Table 3 shows the correctly and incorrectly classified instances for the various methods. Table 4 has classification result analysis, showing Mean absolute error, Root mean squared error, Relative absolute error and Root relative squared error.

## 7    Conclusion

Classification of web services using decision trees and rules have been investigated using various methods like DecisionStump, J48, LMT, Random Forest, RandomTree, REPTree, Decision Table, JRip, OneR, PART and ZeroR in WEKA framework. The above experimental study shows that Decision Tree and JRip algorithms are performing very well with 100% accuracy. But JRip is optimal algorithm with 100% accuracy, no errors and taking minimum amount of time/sec. for classification of quality of web services.

Since decision rule algorithms are simple, easy to understand and require less time for rules generation and can work on multiple attributes, it is advisable to use JRIP algorithm. Future work includes implementation of mechanism for the choice of QoS parameters information when search queries are executed.